\documentclass[twocolumn,prl]{revtex4}

\def \e {\epsilon}
\def \o {\omega}

\begin{document} 

\title{The Boulware-Deser mode in Zwei-Dreibein gravity}

\author{M\'{a}ximo Ba\~{n}ados$^{1} $, Cedric Deffayet$^{2} $ and Miguel Pino$^{3} $  }

 \affiliation{$^{1}$ Instituto de F\'{\i}sica, P. Universidad Cat\'{o}lica de Chile, Casilla 306, Santiago 22, Chile. \\
 $^{2} $UPMC-CNRS, UMR7095, Institut d’Astrophysique de Paris, GReCO, 98bis boulevard Arago, F-75014 Paris, France.\\
 $^{3} $Departamento de F\'isica, Universidad de Santiago de Chile, Av. Ecuador 3493, Santiago, Chile.}

\begin{abstract}
Massive gravity in three dimensions accepts several different formulations. Recently, the 3-dimensional bigravity dRGT model in first order form, Zwei-Dreibein gravity, was considered by  Bergshoeff {\it et al.} and it was argued that the Boulware-Deser mode is killed by extra constraints. We revisit this assertion and conclude that there are sectors on the space of initial conditions, or subsets of the most general such model, where this mode is absent. But, generically, the theory does carry 3 degrees of freedom and thus the Boulware-Deser mode is still active. Our results also sheds light on the equivalence between metric and vierbein formulations of dRGT model. 
\end{abstract}

 \maketitle

The search for a well-defined, unitary, stable, massive version of general relativity has seen huge interest in recent years (for a review see \cite{Hinterbichler:2011tt}). 
de Rham, Gabadadze and Tolley (henceforth dRGT) have recently constructed a theory of massive gravity \cite{deRham:2010ik}, \cite{deRham:2010kj}, \cite{deRham:2011rn},  for which there seems to be an agreement (following in particular the works of \cite{Hassan:2011zd},\cite{Hassan:2011vm})
that the 
Boulware-Deser mode \cite{Boulware:1972zf},\cite{Boulware:1973my} --an instability of non-linear Pauli-Fierz \cite{Fierz:1939ix} theory-- is eliminated by a wise choice of interaction potential.

A particularly simple and nice formulation of dRGT gravity was put forward in  \cite{Hinterbichler:2012cn} (see also \cite{Deffayet:2012nr,Deffayet:2012zc} for a discussion on the equivalence between metric and vielbein formulations). The action is built using vielbeins 1-forms and their corresponding 2-forms curvatures. 
A three dimensional version of this formulation, which can shed light on the four dimensional one, has recently been considered in \cite{Bergshoeff:2013xma}. The action is
 \begin{eqnarray}
I=\int   \left(\hat R_a \hat e^a +  \hat Q_a  \hat \ell^a  +k_1\, \e_{abc} \hat e^a  \hat e^b  \hat \ell^c +k_2\,  \e_{abc} \hat \ell^a  \hat \ell^b  \hat e^c  \right),
\label{action1}
\end{eqnarray}
where $\hat e^a$ and $ \hat \ell^a$ are two independent dreibeins. Here and henceforth wedge product are implicit. The connections are denoted by $\hat w^a$ and $\hat \pi^a$ with curvatures 
\begin{eqnarray}
\hat R^{a}=d \hat\o^{a}-\frac{1}{2}\e^a_{\;\,bc} \hat\o^b \hat\o^c, \ \ \ \  \hat Q^{a}=d \hat\pi^{a}-\frac{1}{2}\e^a_{\;\,bc} \hat\pi^b \hat\pi^c.   \nonumber
\end{eqnarray}
All hatted quantities are spacetime forms. The corresponding spatial forms will be denoted by the same letter without the hat. Latin indexes are raised and lowered with Minkowski metric $\eta^{ab}$ and $\eta_{ab}$.  For simplicity we do not incorporate cosmological constants at each sector. $k_1$ and $k_2$ are free parameters.

It was argued in \cite{Bergshoeff:2013xma} that (\ref{action1}) does not carry a Boulware-Deser mode, in agreement with the 4-dimensional claims (mostly based on the metric formulation, see however \cite{Hinterbichler:2012cn,Deffayet:2012nr,Alexandrov:2013rxa}). The goal of this Letter is to critically analyze this issue. Our conclusion will be that the Boulware-Deser mode 
 is generically still active in the formulation (\ref{action1}) even though there are indeed subcases where it is absent.

The simplicity of working in three dimensions is seen by the fact that the action (\ref{action1}) is already in Hamiltonian form. One only needs to perform a 2+1 decomposition of forms, 
\begin{eqnarray}
\hat e^a_\mu\, dx^\mu = e^a_i dx^i + e^a_0 dt \equiv e^a + e^a_0 dt \\
\hat \ell^a_\mu\, dx^\mu = \ell^a_i dx^i + \ell^a_0 dt \equiv \ell^a + \ell^a_0 dt
\end{eqnarray} 
and likewise for $\hat w^a_\mu dx^\mu$ ,$\hat \pi^a_\mu dx^\mu$. The action in the 2+1 decomposition becomes
\begin{eqnarray}\label{IH}
I&=&\int  \dot{\o}^ae_a + \dot{\pi}^a \ell_a  + \o^a_{\;\,0} De_a +\pi^a_{\;\,0} \nabla \ell_a \nonumber\\
&& +e^c_{\;\,0} \big( R_c+2k_1\e_{abc}e^a\ell^b+k_2\e_{abc}\ell^a\ell^b\big)\nonumber\\
&& +\ell^c_{\;\,0} \big( Q_c+2k_2\e_{abc}\ell^ae^b+k_1\e_{abc}e^ae^b\big),
\end{eqnarray}
where one can read the symplectic structure in a straightforward way. Here `dot' stands for time derivative and  
\begin{eqnarray}
D v^a=d v^a-\e^a_{\;\,bc}\o^bv^c, \ \ \ \ \ \nabla v^a=d v^a- \e^a_{\;\,bc}\pi^bv^c.
\end{eqnarray}
Note that we still use form notation on the 2-dimensional spatial manifold.  

The spatial fields $\{e^a_i,w^b_j\}$ and $\{\ell^a_i,\pi^b_j\}$ form 12 canonical pairs, while the temporal components $e^a_0,\ell^a_0,w^a_0,\pi_0^a$ are 12 Lagrange multipliers. This property is characteristic of generally covariant systems and, as we shall remark below, has important consequences on the consistency algorithm. 

 Let us do a  first counting of degrees of freedom based on the number of canonical variables and constraints (we shall argue below that there are no secondary constraints in the most generic case). There are 24 canonical variables and 12 constraints. The gauge symmetries are 6 (3 overall Lorentz transformations plus 3 overall diffeomorphisms). Thus, among the 12 constraints, 6 of them must be first class, and the remaining 6 must be second class. The Lagrange multipliers do not contribute to the degrees of freedom count.  Each first class constraint kills two canonical variables and each second class constraint kills one. The expected number of degrees of freedom is then
\begin{equation}\label{dof}
{1 \over 2}\left( 24 - 2\times 6 - 6 \right) = 3.
\end{equation} 

There  are two aspects one needs to check to validate this counting. First, we must prove  that are no secondary constraints arising from consistency of the primary ones. Second, we must prove that there are no hidden ``undiscovered" gauge symmetries which could increase the number of first class constraints. 

We start by arguing that a Lagrangian of the form (\ref{IH}) does not give rise to secondary constraints, at least on its generic sector (to be defined below). To simplify the notation, let us collect all canonical variables in pairs $q^i,p_i$, and the Lagrange  multipliers in a set $\lambda^\alpha$. The action (\ref{IH}) has the form 
\begin{equation}\label{H0}
I[q^i,p_j,\lambda^\alpha]=\int dt( p_i \dot q^i - \lambda^\alpha \phi_\alpha(q,p)),
\end{equation} 
where $\phi_\alpha(q,p)$ are functions of the canonical variables and the Lagrange multipliers $\lambda^\alpha$ are independent variables. Variation of this action with respect to $\lambda^\alpha, p_i$ and $q^i$  yields the equations of motion, 
\begin{equation}\label{eqn}
\phi_\alpha(p,q)=0,  \ \ \ \ \ \dot q^i = {\partial \phi_\alpha \over \partial p_i} \lambda^\alpha, \ \ \ \ \ 
\dot p_i = -{\partial \phi_\alpha \over \partial q^i} \lambda^\alpha, 
\end{equation} 
respectively. The constraints $\phi_\alpha=0$ must be preserved in time. Thus, they must satisfy,
\begin{eqnarray}
{d\phi_\alpha \over dt} = {\partial \phi_\alpha \over  \partial q^i} \dot q^{i}   + {\partial \phi_\alpha \over  \partial p_i} \dot p_i = [\phi_\alpha,\phi_\beta]\lambda^\beta \approx 0. \label{consi}  
\end{eqnarray}
In the second equality we have used the equations of motion (\ref{eqn}).

Despite the fact that (\ref{consi}) are algebraic (no time derivatives), they are {\it not} constraints in the Dirac sense because they contain the Lagrange multipliers. When analyzing (\ref{consi}), three different situations may arise: 
\begin{enumerate}

\item First class algebra: If $[\phi_\alpha,\phi_\beta]= f^{\gamma}_{\ \alpha\beta} \phi_\gamma$ for some $f^{\gamma}_{\ \alpha\beta}$, then (\ref{consi}) impose no conditions. The Lagrange multipliers are not fixed by the equations leaving undetermined functions. In this situation, there is a gauge symmetry generated by $\phi_\alpha$.

\item Second class constraints: If $[\phi_\alpha,\phi_\beta]$ is invertible, then Eq. (\ref{consi}) implies $\lambda^\beta =0$. There are no undetermined functions and no gauge symmetry.  

\item Mixed case: If $[\phi_\alpha,\phi_\beta]$ has a some non-zero eigenvalues, then some Lagrange multipliers are fixed and some are arbitrary. There is a gauge symmetry generated by the subset of constraints satisfying a first class algebra. Our system belong to this class. 
\end{enumerate}

In any event, equations (\ref{consi}) either impose no conditions at all or can be solved as restrictions on the Lagrange multipliers, without implying secondary constraints \footnote{If the action has a non-zero Hamiltonian, $H_0(p,q)$, then the consistency algorithm may yield secondary constraints. 
The linearized theory on the maximally symmetric bakground has a non-zero Hamiltonian, further conditions do arise and kill the Boulware-Deser mode.\label{f1} }. 

Having said that, let us note the following possible bifurcations of the consistency algorithm. Generically, the matrix of Poisson bracket of constraints
\begin{equation}
[\phi_\alpha, \phi_\beta] \equiv C_{\alpha\beta}(p,q)
\end{equation} 
is a function of the canonical coordinates $q^i,p_j$. Then, the conditions (\ref{consi}) could be solved, for example, by imposing $C_{\alpha\beta}(p,q)=0$, rather than as a condition on $\lambda^\alpha$. (Of course a mixed case where some components of $C_{\alpha\beta}$ vanish is also possible.) This introduces new constraints on the canonical variables $q,p$. The algorithm has to be run again making sure that $C_{\alpha\beta}(p,q)$ is preserved. Provided the new consistency conditions can be carried to a good end, one has found a different sector of the space of solutions, with further constraints. This branch will carry less degrees of freedom. The Boulware-Deser mode in Zwei-Dreiben gravity, as discussed in \cite{Bergshoeff:2013xma}, was killed in this way, as we will argue. However there exists other branches (in fact more generic) where this  mode {\it is} present.

For any theory, the generic sector is the one where (\ref{consi}) is solved by conditions on the Lagrange multipliers. (In the case of a first class algebra, (\ref{consi}) is automatically satisfied.)  This sector imposes a minimum set of constraints on the initial conditions and carry the maximum number of degrees of freedom. Other sectors, where (\ref{consi}) is solved as conditions on $p,q$ carry less degrees of freedom. These restricted sectors are unstable under generic perturbations of the initial conditions.

Let us derive (\ref{consi}) for the system described by the action (\ref{action1}). It is interesting to note that these equations can be derived without actually computing Poisson brackets (for completeness, we exhibit the Poisson brackets below anyway).  Start from the covariant equations 
\begin{eqnarray}
\hat R_a &=& -\epsilon_{abc}( 2k_1 \hat e^b \hat \ell^c + k_2 \hat \ell^b\hat  \ell^c)  \label{1}\\
\hat Q_a &=& -\epsilon_{abc}( k_1 \hat e^b \hat e^c + 2k_2 \hat e^b \hat \ell^c) \label{2}\\
\hat D\hat e^a &=& 0   \label{3}\\
\hat \nabla \hat \ell^a &=& 0 \label{4}
\end{eqnarray} 
which follow by varying (\ref{action1}) with respect to $e^a,\ell^a,w^a,\pi^a$.
These equations satisfy some integrability conditions. First, the curvatures satisfy the Bianchi identities $\hat D\hat R_a=0$ and $\hat \nabla \hat Q_a=0$  (wedge symbols omitted). Second, the covariant derivatives satisfy Cartan equations $\hat D\hat D\hat e^a=\epsilon^{a}_{\ bc}\hat R^b \hat e^c$ and $\hat \nabla\hat \nabla\hat \ell^a=\epsilon^{a}_{\ bc}\hat Q^b \hat \ell^c$.  Using these relations on equations (\ref{1}-\ref{4}) we derive three algebraic relations, 
\begin{eqnarray}
(k_1 \hat e^a + k_2 \hat \ell^a) \hat e_b\hat  \ell^b\!\!\!\!&=&\!\!\!\! 0 \label{c1}\\
(k_1 \hat e_b + k_2 \hat \ell_b)(\hat w^b-\hat \pi^b) \hat e^a + k_2(\hat w^a-\hat \pi^a)\hat \ell_b \hat e^b \!\!\!\!&=&\!\!\!\! 0 \label{c2}\\
(k_1 \hat e_b + k_2 \hat \ell_b)(\hat w^b-\hat \pi^b) \hat \ell^a + k_1(\hat w^a-\hat \pi^a)\hat e^b \hat\ell_b   \!\!\!\!&=&\!\!\!\! 0 \label{c3}
\end{eqnarray} 
These algebraic relations are {\it not} constraints, in the Dirac sense, because they mix the canonical variables ($ e^a_i,l^a_i...$) with the Lagrange multipliers ($e^a_0,l^a_0...$). In fact, equations (\ref{c1}),(\ref{c2}) and (\ref{c3}) are exactly the consistency relations (\ref{consi}) for this particular theory. To convince oneself that (\ref{c1}),(\ref{c2}) and (\ref{c3}) are linear in the Lagrange multipliers, as (\ref{consi}), it is enough to note that they are 3-form equations and thus each term will contain one, and only one, factor of $e^a_0,l^a_0,w^a_0$ or $\pi^a_0$. 

In Ref. \cite{Bergshoeff:2013xma}, Eq. (\ref{c1}) was solved imposing  
\begin{eqnarray} \label{symvier}
\hat e_a \hat l^a=0.
\end{eqnarray}
Then, (\ref{c2}) and (\ref{c3}), together with invertibility of $\hat{e}^a$ and $\hat{\ell}^a$, imply $(k_1 \hat e_b + k_2 \hat \ell_b)(\hat w^b-\hat \pi^b)$=0. The spatial projections of these equations are, 
\begin{equation}\label{eilj}
\chi_1\equiv e_{a} l^{a} =0, \ \ \ \ \  \chi_2\equiv (k_1 e_{b} + k_2 \ell_{b})( w^b- \pi^b)=0.\nonumber
\end{equation} 
These equations are secondary constraints in the Dirac sense, i.e., algebraic relations involving only the canonical variables (not the Lagrange multipliers). However, the complete equations are (\ref{c1}),(\ref{c2}) and (\ref{c3}) not (\ref{eilj}). One can find particular solutions satisfying  (\ref{eilj}) but they do not capture the whole theory. Indeed, there exist solutions satisfying (\ref{c1}),(\ref{c2}),(\ref{c3}) but not (\ref{eilj}) (or (\ref{symvier})) .
An example is provided by the fields, 
\begin{eqnarray}
\hat{e}^{a}_{\ \mu} &=& \left( \begin{array}{ccc}
r & 0 & 0 \\ 
0 & { \sqrt{k_2} \over \sqrt{2} k_1} {1 \over r}  & 0 \\ 
0 & 0 & r
\end{array} \right), \label{eos} \\
\hat{\ell}^{a}_{\ \mu} &=& \left( \begin{array}{ccc}
-{k_1  \over k_2} r & 0 & 0 \\ 
0 & {c \over r} & {r\over k_2}\sqrt{2c^2k_1k_2^2-k_1^2 } \\ 
0 & 0 & -{k_1 \over k_2}r
\end{array}   \right) \label{los}
\end{eqnarray} 
where $c$ is an integration constant. These fields are perfectly reasonable solutions ($\hat{e}^a$  describes anti-de Sitter space) and the combination 
\begin{equation}
 \hat{e}_{a}\hat{\ell}^{a} = - {\sqrt{ 2c^2k_1 k_2^2 - k_1^2} \over \sqrt{2k_2} k_1 }\, drd\phi \neq 0
\end{equation} 
is not zero.  These fields can easily be generalized to a solution where $\chi_2$ is also non-zero. 

There are however cases where (\ref{c1}),(\ref{c2}) and (\ref{c3}) do lead to extra constraints. For example, if $k_1$=0 ($k_2$=0) then invertibility of $\ell^a$ ($e^a$) implies that  (\ref{eilj}) must hold. More generally, on the subset of solutions where $k_1 \hat e^a + k_2 \hat \ell^a$ is invertible as a matrix, equations (\ref{c1}) imply that (\ref{eilj}) must hold. However,  
invertibility of $k_1 \hat e^a + k_2 \hat \ell^a$ does not follow from action (\ref{action1}). For example, $k_1 \hat e^a + k_2 \hat \ell^a$ evaluated on the solution (\ref{eos}),(\ref{los}) is not invertible, even though the vierbeins $\hat e^a$ and $\hat \ell^a$ are. Insisting upon invertibility of $k_1 \hat e^a + k_2 \hat \ell^a$ has to be imposed as an extra constraint on the theory. Note that this is fully consistent with the analysis of \cite{Deffayet:2012nr} where it was shown (in the 4D case) that the field equation were not leading in the generic case to equation (\ref{symvier}) -- in contrast to what is said e.g. in \cite{Hinterbichler:2012cn}-- while on the other hand, the condition (\ref{symvier}) was entering in a crucial way in the elimination of the BD ghost. Note that condition (\ref{symvier}) plays also a crucial role in showing the equivalence between the vielbein and metric formulation of dRGT theory. Indeed, the later formulation involves a matrix square root which was shown in \cite{Deffayet:2012zc} to exist iff (in the three and four dimensional cases) vielbeins obeying condition (\ref{symvier}) can be chosen.

Let us now go back to a canonical language and check our statement  
that the action (\ref{action1}) does not have any hidden `undiscovered" gauge symmetries. To this end, we now compute the brackets of all constraints with themselves and extract the number of first and second class ones.  
It is easier to consider the smeared constraints,
\begin{eqnarray}
\Phi_1(\xi)\!\!\!\!&=&\!\!\!\!\int\xi_a De^a\nonumber\\
\Phi_2(\chi)\!\!\!\!&=&\!\!\!\!\int\chi_a\nabla l^a\nonumber\\
\Phi_3(\xi)\!\!\!\!&=&\!\!\!\!\int \xi^c \big(R_c+\e_{abc}\big[2k_1e^al^b+k_2l^al^b\big]\big)\nonumber\\
\Phi_4(\chi)\!\!\!\!&=&\!\!\!\!\int\chi^c\big(Q_c+\e_{abc}\big[2k_2l^ae^b+k_1e^ae^b]\big)\nonumber
\end{eqnarray}    
Here, $\xi$ and $\chi$ are arbitrary functions on the spatial manifold, which will be removed at the end. Note that the test functions are vectors; each functional $\Phi_i$ carries three constraints. 

The Poisson brackets of constraints are found easily as
\begin{eqnarray}
\left[\Phi_1(\xi),\Phi_1(\chi)\right]\!\!\!\!&=&\!\!\!\! \int -\e^{ab}_{\;\;\;c}\xi_a\chi_b De^c\nonumber\\
\left[\Phi_1(\xi),\Phi_2(\chi)\right]\!\!\!\!&=&\!\!\!\! 0\nonumber\\
\left[\Phi_1(\xi),\Phi_3(\chi)\right]\!\!\!\!&=&\!\!\!\!\int 2k_1\e_{dbc}\e^{d}_{\;\;ae}\chi^b\xi^a \ell^c e^e -\e_{abc}\xi^a\chi^b R^c  \nonumber\\
\left[\Phi_1(\xi),\Phi_4(\chi)\right]\!\!\!\!&=&\!\!\!\! \int2\e_{dbc}\e^{d}_{\;\;ae}\chi^b\xi^a\big(k_1e^c+k_2\ell^c\big)e^e \nonumber\\
\left[\Phi_2(\xi),\Phi_2(\chi)\right]\!\!\!\!&=&\!\!\!\!\int -\e^{ab}_{\;\;\;c}\xi_a\chi_b \nabla l^c\nonumber\\
\left[\Phi_2(\xi),\Phi_3(\chi)\right]\!\!\!\!&=&\!\!\!\! \int2\e_{dbc}\e^{d}_{\;\;ae}\chi^b\xi^a\big(k_1e^c+k_2\ell^c\big)\ell^e \nonumber\\
\left[\Phi_2(\xi),\Phi_4(\chi)\right]\!\!\!\!&=&\!\!\!\! \int 2k_2\e_{dbc}\e^{d}_{\;\;ae}\chi^b\xi^a e^c \ell^e - \e_{abc}\xi^a\chi^b Q^c \nonumber\\
\left[\Phi_3(\xi),\Phi_3(\chi)\right]\!\!\!\!&=&\!\!\!\! \int2k_1 \e_{abc}\xi^a\chi^bD\ell^c\nonumber\\
\left[\Phi_3(\xi),\Phi_4(\chi)\right]\!\!\!\!&=&\!\!\!\ \int-2\e_{abc}\big(D\xi^a \chi^b+\xi^a \nabla \chi^b \big) \big(k_1e^c+k_2l^c \big)
\nonumber\\
\left[\Phi_4(\xi),\Phi_4(\chi)\right]\!\!\!\!&=&\!\!\!\ \int2k_2\e_{abc}\xi^a\chi^b\nabla e^c\label{pb}
\end{eqnarray}
To justify the counting (\ref{dof}) we should diagonalize (\ref{pb}) isolating first and second class constraints to find six of each. But this is not so easy and we shall not  
attempt to do it here \footnote{The generator of overall Lorentz transformations can be easily isolated but it is not necessary. }. Instead, we ask the restricted question of how many zero and how many non-zero eigenvalues does the matrix (\ref{pb}) have, when evaluated on the constraint surface, i.e. what is the rank of the matrix of constraints $[\Phi_\alpha,\Phi_\beta]$.

This question does not have a unique answer because the right hand side of (\ref{pb}) is field dependent. For example, the proportional ansatz \cite{Bergshoeff:2013xma} $\ell^a =\alpha e^a$ yields solutions with rank=4, while the family of solutions (\ref{eos}) and (\ref{los}) have the maximum rank=6. The maximum rank is 6 because the action (\ref{action1}) has 6 gauge symmetries and twelve constraints. Thus, at least 6 constraints are first class. 

The sector with maximum rank is crucial because is stable under perturbations of initial conditions. On open sets around maximum rank solutions one can forget about the field-dependent nature of $[\phi_\alpha,\phi_\beta]$; The number of zero and non-zero eigenvalues becomes meaningful representing, respectively, the number of first and second class constraints. A similar situation arises in higher dimensional Chern-Simons theories \cite{Banados:1995mq}.

Since the maximum rank can be achieved, for example by (\ref{eos}) and (\ref{los}), we conclude that on its generic sector this theory has 6 first class constraints and 6 second class ones. The counting (\ref{dof}) is thus correct and the Boulware-Deser mode is active.

Summarizing, Zwei-Dreibein gravity does have an active Boulware-Deser mode. There exists sectors in the space of initial conditions and also a subset of theories where this mode is hidden, but these are not generic. We emphasize that we have not uncover in this paper the nature of this mode. While it is expected to be a ghost, this conclusion needs explicit confirmation.

A final word on the applications of this result to 4 dimensions is in order. A detailed Hamiltonian analysis in 4 dimensions seems to confirm the elimination of the Boulware-Deser mode (see \cite{Alexandrov:2013rxa}, and references therein). See \cite{Banados:1997hs} for an alternative Hamiltonian formulation of vierbein gravity. However, to our knowledge, the issue of bifurcations and maximum rank conditions has not been analyzed in detail. The 4-dimensional calculation is further complicated by several issues discussed in \cite{Henneaux:1983vi},\cite{Charap:1988vz}. We hope to come back to this important case in the near future. Note in particular that it was noticed in the Lagrangian analysis of \cite{Deffayet:2012nr} using the vielbein formulation, that the way the BD ghost was removed was very different depending on the mass term considered (i.e. translating in the present formalism, depending on the vanishing or non vanishing of $k_1$ and/or $k_2$). In one case, the ghost was argued to be absent {\it per se}, while in other cases, its removal required the {\it extraneous} assumption of a constraint of the form (\ref{symvier}), while in some other cases even such an assumption was not enough to conclude. This is in complete argreement with the present analysis, but also opens the possibility that in 4 dimensions as well the BD ghost might still be present in the most general case (at least in the vierbein formulation), while it could be absent in a subset of theories.

~ 

~ 

We would like to thank E.~A.~Bergshoef, S.~de Haan, O.~Hohm, W.~Merbis and P.K.~Townsend for important discussions on the first version of this Letter, and also W. Merbis for correcting an important misprint. Useful remarks by S. Carlip, J. Gomis and M. Henneaux are 
also acknowledged. MB is partially supported by Fondecyt (Chile) \#1100282 and Anillo ACT (Chile)  \#1102. MP is partially funded by Conicyt, project 7912010045. The research leading to these results has received funding from the European Research Council under the European Community’s Seventh Framework Programme (FP7/2007-2013 Grant Agreement no. 307934). CD and MB thank the APC lab in Paris, where this research was initiated.
     
     ~
        

\end{document}